\documentstyle[11pt,aaspp4]{article}

\def\lta{\ifmmode {\,\mathbin{\lower 3pt\hbox   
    {$\,\rlap{\raise 5pt\hbox{$\char'074$}}\mathchar"7218\,$}}}
    \else {${\mathbin{\lower 3pt\hbox
    {$\rlap{\raise 5pt\hbox{$\char'074$}}\mathchar"7218\,$}}}
    $}\fi}
\def\gta{\ifmmode {\mathbin{\lower 3pt\hbox   
    {$\,\rlap{\raise 5pt\hbox{$\char'076$}}\mathchar"7218\,$}}}
    \else {${\mathbin{\lower 3pt\hbox
    {$\rlap{\raise 5pt\hbox{$\char'076$}}\mathchar"7218\,$}}}
    $}\fi}

\def\etal{et\ al.}

\begin{document}

\lefthead{HOLZ, MILLER, \& QUASHNOCK}
\righthead{GRAVITATIONAL LENSING LIMITS ON GRB REDSHIFTS}

\title{Gravitational Lensing Limits on the Average Redshift of Gamma-Ray Bursts}

\author{Daniel E.\ Holz}
\affil{Department of Physics, University of Chicago\\
       5640 South Ellis Avenue, Chicago, IL 60637-1433\\
       deholz@rainbow.uchicago.edu}
\authoremail{deholz@rainbow.uchicago.edu}

\author{M.\ Coleman Miller}
\affil{Department of Astronomy and Astrophysics, University of Chicago\\
       5640 South Ellis Avenue, Chicago, IL 60637-1433\\
       miller@bayes.uchicago.edu}
\authoremail{miller@bayes.uchicago.edu}

\author{Jean M.\ Quashnock}
\affil{Department of Astronomy and Astrophysics, University of Chicago\\
       5640 South Ellis Avenue, Chicago, IL 60637-1433\\
       jmq@oddjob.uchicago.edu}
\authoremail{jmq@oddjob.uchicago.edu}

\slugcomment{To appear in {\em The Astrophysical Journal}}

\received{1998 April 28}
\revised{1998 August 3}
\accepted{1998 August 6}

\begin{abstract}

The lack of bright host galaxies in several recently
examined gamma--ray burst (GRB) error boxes suggests
that the
redshifts of cosmological GRBs may be significantly
higher than previously hypothesized.
On the other hand,
the non--detection of multiple images in the 
BATSE 4B catalog implies an upper limit 
to the average redshift $\langle z\rangle$ of GRBs.
Here, we calculate an upper limit to $\langle z\rangle$, 
independent of the physical model for GRBs, 
using a new statistical lensing method that removes
distance ambiguities, and thus permits accurate computation
of the lensing rate at high $z$.
The upper limit on $\langle z\rangle$ depends 
directly on the cosmological parameters $\Omega$ and $\Lambda$.
If there are no multiple images among the brightest 80\%
of the first 1802 bursts in the
BATSE 4B catalog, then, at the 95\% confidence level,
$\langle z\rangle<$ 2.2, 2.8, 4.3, or 5.3 for ($\Omega$, $\Lambda$)
values of 
(0.3, 0.7), (0.5, 0.5), (0.5, 0.0), or (1.0, 0.0), respectively.
The 68\% upper limit to the
average redshift is comparable to or less than the median redshift
of GRBs in scenarios in which the GRB rate is proportional
to the rate of star formation, for any cosmology.
The uncertainty in the lensing rate---arising
from uncertainties in the cosmological parameters 
and in the number density and average velocity dispersion of
galaxies---will be reduced significantly in the next few years
by a new generation of experiments and surveys.
Moreover, the continued increase
in the number of GRBs observed by BATSE will greatly
constrain their redshift distribution.

\end{abstract}


\section{INTRODUCTION}

Three decades after their discovery (\cite{KSO73}),
the physical origin of gamma--ray bursts (GRBs) remains unresolved.
Recent developments, however---including the isotropy
of GRBs seen by BATSE (\cite{B96}; \cite{THBM96}) 
and the detection of redshift
$z$=0.835 absorption and emission lines from a possible optical
counterpart to GRB 970508 (\cite{M97}; \cite{R98})---strongly
suggest a cosmological origin for GRBs.

Initially, no--evolution fits to the log N -- log P (peak flux) 
distribution of BATSE GRBs suggested a typical redshift of
$z\sim 1$ for dim bursts (\cite{F93}; \cite{W93}),
with the break in the $-3/2$~slope of the flux distribution
at $P \sim 10$ ph~cm$^{-2}$~s$^{-1}$ being interpreted as 
a cosmological deviation from Euclidean space at $z \sim 1$. 
In addition, a number of researchers reported a factor of $\sim 2$ 
``time--stretching'' of
dim bursts relative to bright ones (e.g., \cite{Nor94}, 1995), 
which was thought to imply a redshift of order unity for the 
dim bursts. Such a redshift would imply an extremely low rate
of gravitational lensing and multiple imaging of sources detected with
BATSE, perhaps as low as one multiple imaging event per 200
years (\cite{GN94}).

New lines of evidence
now indicate that the typical redshift of cosmological GRBs
may substantially exceed unity.
Particularly suggestive 
evidence for high GRB redshifts comes from 
deep HST searches of the error boxes of several of the brightest GRBs
detected by BATSE and the Interplanetary network (IPN):
In five cases, no obvious host galaxies
for the bursts were found, down to a limiting magnitude
between 3.5 and 5.5 mag fainter than what would be expected
if GRBs reside in $L_\star$ galaxies and dim bursts are
at $z\sim 1$ (\cite{Sch97}).
This ``no--host problem'' has been analyzed further by
Band \& Hartmann (1998), and implies
that if GRBs reside in normal galaxies,
even the {\em brightest} of them are at redshifts close to unity,
with the faintest being at much higher redshift still.
In addition, Fenimore \& Bloom (1995) showed that the intrinsic 
anticorrelation between photon energy and burst duration implies 
that if the observed time--stretching is caused by time dilation, then
the redshift of the dimmest bursts may be as large as
$z\sim 6$ (note,
however, that some or all of the time--stretching may be
intrinsic to the bursts; cf. \cite{SPS97}).

These new lines of evidence have led researchers to explore
scenarios in which the bursts
are at much higher redshifts. One currently popular scenario,
motivated in part by the assumption that cosmological
GRBs involve remnants of massive stars, is that the
GRB rate is proportional to the rate of (massive) star formation
(\cite{Tot97}; \cite{WBBN98}). The star formation rate
is thought to vary strongly with redshift (\cite{Ma96}),
peaking at $z \sim 2$.
If this is the case, the dimmest bursts may be at redshift
$z\sim 6$, making them the most distant objects
ever detected (\cite{WBBN98}). Other authors have noted that,
if the comoving number density of GRB sources is allowed to
vary with redshift, then even if the peak luminosity is
fixed, the log N--log P distribution and other properties of
the BATSE bursts can be fit by models with a maximum redshift
up to $z\sim$10--200 (\cite{RHL95}).

In these high-$z$ scenarios, the expected incidence of gravitational lensing 
and multiple imaging of GRB sources is much higher than it is in
lower-$z$ scenarios,
since in general the lensing rate increases markedly with the 
source redshift.  The short duration of GRBs (typically tens of seconds) 
compared to the difference in
light travel time between different ray paths 
(typically months, if the lens is of galactic mass; cf. \cite{M92}) 
means that a multiply-imaged GRB  
appears as two or more separate events
with identical time histories and intensities that differ only by
a scale factor. The image separations, of order arcseconds,
are tiny compared to BATSE location errors; thus, these GRBs would
appear to come from the same location.

Two or more such events must be detected in order to identify
a lensed GRB source; thus, the overall BATSE burst detection efficiency,
$\epsilon$, is of prime importance
in determining the observed incidence of lensing.
The average efficiency for the 4B catalog is $\epsilon = 0.48$
(\cite{M98}), which is 40\% larger than the previously estimated
value (\cite{F94}). This increase implies that the expected incidence
of lensing is significantly higher than was previously believed.

Lensing of GRBs was first suggested by Paczynski (1986) 
as a way to establish their cosmological origin.  
Subsequently, many authors have calculated the lensing rate of GRBs 
(\cite{M92}; \cite{BW92}; \cite{N93}; \cite{GN94}),  
assuming GRB redshifts estimated from no--evolution fits to the
log N -- log P distribution, viz., $z\lta 1$.
For redshifts $z \lta 1$, the gravitational lensing rate
is reasonably well known for a given cosmology (e.g., \cite{TOG84};
\cite{FT91}; \cite{F92}).
At redshifts $z \gta 1$, however, the lensing rate has been uncertain, 
mainly because of ambiguity in the angular diameter distance at high redshift 
(\cite{F92}).  As a result, even when the cosmology and the 
number density and properties of lenses are fixed, the estimated
lensing rate can vary by factors of several between the different
prescriptions.

Motivated in part by this ambiguity, Holz \& Wald (1998) have
developed a numerical method to calculate lensing
rates for a given cosmology, combining
techniques from both ``ray-shooting'' and ``Swiss-cheese'' models.
The approach resolves the angular-diameter distance uncertainties,
and also correctly accounts for multiple lens encounters
along the line of sight, allowing for an unambiguous calculation
of lensing rates.

If there are no multiple images among the bursts 
detected with BATSE, then an upper limit 
to the average redshift $\langle z\rangle$ of GRBs can be inferred,
one which depends directly on the cosmological parameters
(see also Nemiroff et al.\ 1994; Marani et al.\ 1998;
Marani 1998).
Here, we set an upper limit to $\langle z\rangle$, 
independent of the physical model for GRBs, 
using the Holz \&~Wald (1998) method
to compute the lensing rate for a variety of cosmologies.

The plan of this paper is as follows.
In \S~2 we describe and develop the statistical lensing method, 
and compare its results with analytical estimates. We show
that analytical estimates using the ``filled-beam" approach
are reasonably accurate for source redshifts less than
$\sim$2--3, but underestimate the true rate of lensing
by tens of percent at higher redshifts. 
In \S~3 we use the numerical method to compute the lensing rate 
as a function of redshift for several cosmologies, 
and compare these results with previous estimates. 
We also  determine upper limits on the rate of lensing 
and on the average redshift $\langle z\rangle$ of GRBs,
assuming that no lensing events are present in the BATSE 4B catalog
(\cite{M98}).  
We discuss the implications of these results
and give our conclusions in \S~4. 
We follow the conventions that the Hubble constant
is 100\,$h$\ km~s$^{-1}$~Mpc$^{-1}$, $\Omega$ is the
present mean density in the universe in units of the closure density,
and $\Lambda$ is the present normalized cosmological constant.
In a flat universe, $\Omega+\Lambda=1$.

\section{GRAVITATIONAL LENSING RATE CALCULATIONS}

We utilize a recently developed numerical model~(\cite{HW98})
to calculate the lensing rate. This method offers a number of distinct
advantages over previous lensing treatments.
The model is fully described and discussed in Holz \& Wald (1998), 
and we refer the reader to that paper for details. 
In \S~2.1 we give an overview of the features of
this method relevant to the calculations reported here,
and in \S~2.2 we compare the numerical results to
analytic lensing rates and lens redshift distributions.

\subsection{Overview of Statistical Lensing Method}

In current analytic methods (see, e.g., Turner \etal\ 1984;~\cite{F92}),
every lensing event is treated as
a photon beam traveling through a universe described by a given
angular diameter
distance expression, and encountering a single lens at some point
along its path.
These methods require distance measures
to calculate lensing probabilities. Common angular diameter
distance measures include the
Dyer-Roeder empty- and filled-beam expressions~(Dyer \&~Roeder 1972, 1973),
and the Ehlers \& Schneider (1986)
variant of these expressions.
The empty-beam distance assumes
that the photon beams encounter no mass aside from that of the lenses,
and otherwise travel through vacuum.
The filled-beam distance presupposes that, aside from the lenses,
the beams encounter exactly the
Robertson-Walker mass density at every point along their trajectories.
These Dyer-Roeder expressions assume that the rays travel far
from all clumps of matter. The Ehlers-Schneider expression attempts
to take into account the statistical effects of a clumpy universe.
At higher redshifts ($z >1$), the differences in lensing rates
inferred from the different distance measures can be
large, and there is no natural way to select between the differing results.
A further assumption in most analytic treatments is that the lensing
is dominated by a single encounter~(Turner \etal\ 1984).
It has been 
unclear to what extent this is a reasonable assumption, 
especially at higher redshifts
where the probability of multiple close encounters becomes large.

For these reasons it is necessary to look at numerically determined
lensing rates, which in general include multiple lens encounters,
and, by integrating along ray-paths in model universes, are able to
eliminate the need for averaged distance expressions.
In particular, the method we use resolves ambiguities
by calculating the distances of a given photon beam directly, bypassing
the need for an averaging expression. In addition, our method
considers every encounter along the line of sight, with contributions
from thousands of possible lenses for a given source.

The model we utilize can be thought of as a combination of
Swiss-cheese models~(\cite{K68})
and ray shooting models~(\cite{PMM98};~\cite{WCO98}).
As in the Swiss-cheese
approach, we decompose the universe into comoving spherical
regions, and allow for mass inhomogeneities
within each region. Unlike most Swiss-cheese treatments, however,
we do not require an opaque core radius, and we permit arbitrary
mass distributions within the comoving spheres.
As is done in ray-shooting approaches, we develop statistics
by considering many random rays.
However, by appropriately treating every individual region along the line
of sight, we avoid the need to collapse the matter near the beam
into lensing planes, as is commonly done in current ray-shooting approaches.

Rather than constructing an image out of many different photons,
we calculate the effects for an infinitesimal bundle of photons
(commonly called a {\em beam}). Instead of using the geodesic equation
for each individual photon component of an image, we use the geodesic deviation
equation for the beam as a whole, and track the lensing parameters
for a given image along a single null geodesic.
We compute the area, shear, and rotation of each beam 
as a function of redshift.
Because these beams represent random directions on the sky,
a large number of such beams constitutes a sample of the past light
cone of an observer.

For our purposes, the {\em area} of the beam 
is the quantity of greatest interest,
as it is inversely proportional to the magnification of the
corresponding image. On occasion 
the area goes to zero at some point along a beam's path (and
then becomes negative, representing an inversion of the image).
This is a {\em caustic}, and once this has happened the
photon beam represents a secondary image.
For every secondary image, there exists
another beam that connects the source and observer without a caustic.
This latter image is the primary image, and one of these exists
for {\em every} source.
The probability that a randomly selected source is multiply imaged is thus
related to the probability that a randomly selected photon beam
has a caustic. 
Most lensing events are dominated by a single encounter with
a lens, and in these cases pairs of images are produced.
If one assumes that all multiple images are pairs (see \S\S\ 3 and 4 below), 
then, by determining the fraction of beams which have a caustic,
it is possible to compute
the probability for multiple imaging, and hence the lensing rate.

The model describes the universe, on {\em all} scales,
by a Robertson-Walker metric that is
perturbed by a weak Newtonian potential $\phi$:
\begin{eqnarray}
\label{metric}
ds^2&=&-(1+2\phi)d\tau^2\\ \nonumber
&&{}+(1-2\phi)a^2\left[{dr^2\over1-kr^2}
+r^2(d\theta^2+\sin^2\theta\,d\varphi^2)\right],
\end{eqnarray}
where $\tau$, $r$, $\theta$, and $\varphi$ are the usual Robertson-Walker
coordinates, $a(\tau)$ is the cosmological expansion
factor, and the curvature of the spatial part of the metric
is either open ($k=-1$), flat ($k=0$),
or closed ($k=1$). We assume that 
the Newtonian potential $\phi$ is small (weak-field), 
that accelerations are non-relativistic,
and that linear curvature terms dominate the non-linear ones: 
\begin{eqnarray*}
\left|\phi\right|&\ll&1\\ 
\left|\partial\phi/\partial\tau\right|^2&\ll&D^a\phi D_a\phi\\
(D^a\phi D_a\phi)^2&\ll&D^b D^c\phi D_b D_c\phi,
\end{eqnarray*}
where $D_a$ is the spatial derivative operator of the metric of
equation~(1),
and where the summation is over all spacetime indices.

As shown in Holz \& Wald (1998),
the Einstein equations for this
metric are globally described by the standard
evolution equations for Robertson-Walker cosmologies.
Locally, however, this metric is equivalent to a
Minkowski spacetime that is weakly perturbed by a Newtonian potential.
As we are solely interested in the statistical distributions of 
photon beams (and, in particular, in scenarios where strong lensing is rare), 
we do not need to consider correlations between photon beams. 
The matter distribution
in the universe far from a photon's trajectory is not
relevant to the lensing, and need not even be known (\cite{HW98}).
We therefore restrict our consideration to matter
in the immediate vicinity of the beam.
With this in mind,
we choose a maximum clustering scale, $\cal R$, for the cosmological
matter. For the purposes of
this paper we have chosen ${\cal R}=2\ \mbox{Mpc}$,
to correspond to typical galaxy separations. 
We divide the mass of the universe into spheres of comoving
radius $\cal R$, each representing an individual galaxy.
These spheres contain a certain amount of Robertson-Walker
matter, with a given distribution for the matter within each sphere.
As discussed in~Holz \&~Wald (1998), clustering on
larger scales does not significantly affect the rate of
lensing by galaxies\footnote{Press \& Gunn~(1973) showed that
the lensing rate from point masses (and, in general, from objects
with mass entirely within their Einstein radii) 
is a function solely of the total mass in the point masses, 
and is independent of their mass distribution.
On the other hand,
if the matter is not confined to within its Einstein radius, then
only the Ricci contribution to lensing matters. This contribution
is completely local and is thus independent of clustering on scales
much less than the observer-source distance.
We thus expect our results to be independent of the value of ${\cal R}$,
and to depend only on the total amount of mass in lenses.
Indeed, we find that our results with ${\cal R}=4\ \mbox{Mpc}$ are
the same as those with ${\cal R}=2\ \mbox{Mpc}$.}
(clustering does produce lensing by larger masses, typically
$10^{14}$--$10^{15}M_\odot$, but the time delay between images
for such lensing is longer than the few year baseline of 
BATSE observations, and hence we neglect it in our analysis).
We have restricted ourselves to distributions in the form of
truncated singular isothermal spheres.
Although we could allow these to evolve in time, in what follows
we have taken the mass distributions to be static;
we do not consider evolution of the lenses (see \S\ 4).

The path of a photon beam in this model universe
can be broken into many steps, 
with lensing effects due solely to the nearest
galaxy calculated at each step.
The distance a photon traverses at any particular step is determined
by the impact parameter with which the photon enters the appropriate
sphere (representing the local galaxy neighborhood).
At no point do we need to arbitrarily specify an average distance
measure, as the distance for each photon beam
is calculated individually.
We consider a photon beam that starts at the observer ($z=0$),
and trace its path backwards in time. 
It enters the first sphere (representing
the first galaxy encountered) of radius
$\cal R$, with a random impact parameter. 
We calculate the lensing of
the beam due to the mass distribution within the sphere.
After traversing the sphere, the photon
proceeds to the next galaxy along its path, again with a random
impact parameter.
As it traverses the boundaries
between spheres, the effects of the cosmological contraction
and blueshift (we are heading backwards in time) are
taken into account. This process is continued from one galaxy
to the next, yielding the lensing parameters
of the photon beam as a function of redshift.
We repeat this process for many different photons, in each case generating
random impact parameters for every galaxy encounter.
Each photon beam is thus completely
uncorrelated with all others, and can be thought to randomly
sample a distinct direction on the sky.
By this Monte-Carlo tracing of photon beams, 
lensing rates can be calculated for a given cosmology.
In this paper a typical simulation will involve tracing the paths
of ${}\sim\!10^4$ independent photon beams.

A detailed discussion of the lensing rate results from the above method, 
including a comparison with results commonly found in the literature, 
will be reported elsewhere. 
In the following subsection we
present results for a number of cosmologies of interest,
and discuss their main features.

\subsection{Comparison of Numerical Results with Analytical Forms}

We compare the lensing probability and
distribution of lens redshifts given by the numerical method
described above with the filled-beam values, for several different
cosmologies. Figure~1 shows the lensing rates for $\Omega$=0.3,
$\Lambda$=0.7; $\Omega$=0.5, $\Lambda$=0.5; $\Omega$=0.5,
$\Lambda$=0; and $\Omega$=1, $\Lambda$=0.
These curves are constructed assuming that the dimensionless
parameter $F\equiv 16\pi^3n_0R_0^3(\sigma_\parallel/c)^4$ is equal
to 0.1, where $n_0$ is the comoving number density of galaxies,
$R_0=c/H_0$, and $\sigma_\parallel$ is the velocity dispersion
inside the galaxies.

A striking feature of the data displayed in Figure~1 is the good
agreement, for redshifts $z\lta3$, between the numerical results and
the results obtained from the analytic filled-beam expressions.  At
redshifts much less than unity, the difference between filled- and 
empty-beam results are slight; all distance measures yield similar
lensing rates. Hence, agreement in this regime is expected.
But at higher redshifts, such a close agreement is not expected 
{\em a priori}.  Why is it obtained?
  
We note that a small $F$ parameter corresponds to galaxies extending 
well outside their Einstein radii.  Matter outside the Einstein radius
contributes to lensing solely through its Ricci contribution to the
convergence of a beam.  This is a local effect, and thus in this case
the lensing of a beam depends upon the amount of matter that the beam
traverses (see, e.g., Holz \&~Wald 1998).  For $F=0.1$, $\Omega=0.5$,
and $\Lambda=0$,
for example, isothermal galaxies have halo
radii of $\approx200\ \mbox{kpc}$.  The Einstein radii of these
objects is $\approx 2{\rm -}5\ \mbox{kpc}$, depending on the
redshifts of the source and lens, so roughly 98\% of
the matter contributes to lensing solely through Ricci effects;
hence the concordance between our numerical results and
the analytic filled-beam expressions.

At much higher redshifts ($z\gta3$), the numerical results begin to
deviate from the filled-beam expressions, yielding larger lensing rates.
At these high redshifts the incidence of multiple lens
encounters becomes significant (Press \&~Gunn 1973; Pei 1993).
For lower redshifts, lensing encounters are invariably dominated by a
single lens, because the probability of multiple close lens encounters is
negligible. However, at higher redshifts multiple close passes
become more common, and these are expected to {\em increase} the
overall lensing rate. Consider, for example, an encounter in which a beam
skims just outside the Einstein radius of a lens, picking up a significant
amount of convergence, but not quite enough to generate a caustic. 
If this beam then comes close to another lens, it is possible for the
additional convergence from this second lens to cause the beam
to caustic.  Although neither of the lens encounters would have yielded 
multiple imaging by themselves, their combined contribution can
cause a secondary image to appear. Thus, the presence of multiple
lenses is expected to increase the lensing rate. A thorough analysis
of the contribution of multiple lenses to lensing will be reported
elsewhere.

Figure~2 shows the distribution of lens redshifts in a universe
with $\Omega$=0.3 and $\Lambda$=0.7, for a source at $z$=3 (panel a)
and for a source at $z$=5 (panel b),
as well as for $\Omega$=1 and $\Lambda$=0 (panels c and d).
The histograms give the
numerically determined distribution, and the solid lines give
the analytical form in the filled-beam approximation.
Again the numerical answers agree reasonably
well with the analytical estimates.
We note that in all cases the bulk of the lenses are at $z\sim 1$.

\section{UPPER LIMITS ON GRB REDSHIFTS FROM LACK OF LENSING}

\subsection{Calculation of the expected number of lensed events}

In \S\ 2 we used the statistical lensing method
(\cite{HW98}) to compute the probability $\tau(z)$ that a point source
at redshift $z$ is multiply imaged. This quantity is dependent 
on the cosmological parameters, $\Omega$ and $\Lambda$, 
as well as parameters describing the model for the lenses.
We have taken for such a model the {\em singular isothermal sphere} (SIS)
model (see, e.g., Turner \etal\ 1984; \cite{FT91}; \cite{K93a}),
wherein the lens is the spherically symmetric 
gravitational potential of a massive dark matter halo of constant temperature.
The lensing probability $\tau(z)$ then depends directly on the parameter
$F$, defined in \S~2.
For realistic models, $0.05 < F < 0.15$; for definiteness,
we take $F=0.1$ (\S 4, see below).

In the SIS model, the lensing event always consists of two
images---the third central image is far too faint to be
observed---with a time delay $\Delta t$ between the two images of order months
for lenses of galactic mass
(\cite{M92}; \cite{GN94}). Thus (\S 4) we  take a constant BATSE efficiency
$\epsilon = 0.48 $ (\cite{M98}) to see either image,
and $\epsilon^2$ to see {\em both}.

Let $N_{\rm tot}$ be the total number of {\em observed} bursts:
For the 5.5 yr observing period of the BATSE 4B catalog (\cite{M98}),
$N_{\rm tot} = 1802$. There are then approximately
$N_{\rm tot}/\epsilon$ {\em actual} 
burst sources above the BATSE threshold during this time. If these sources
are all at a given redshift $z$, the expected number $N$ of image pairs is:

\begin{equation}
	N = \left(N_{\rm tot}/\epsilon\right) \cdot \epsilon^2\cdot \tau(z)
	  = \epsilon N_{\rm tot} \tau(z) \; .
\end{equation}

Of course, not all the burst sources are at the same redshift.
If instead the burst sources distribution in redshift is given by $C(z)$,
where $C(z)$ is the fraction of burst sources with redshift less than $z$,
we then have:

\begin{equation}
	N = \epsilon N_{\rm tot} \int_0^\infty dC(z) \, \tau(z) \; .
\end{equation} 
Furthermore, not all lensed images are actually observable:
In some cases the images will be too faint
to be detected by BATSE (\cite{GN94}). 
We must include the effect of image magnification (or de-magnification)
on the expected number of lenses in equation (3).

We define $b_{\rm min}$ as the brightness threshold below which
identification of burst lenses from lightcurve comparison is impossible.
(Throughout this paper, we measure brightness in units of peak flux
[ph~cm$^{-2}$~s$^{-1}$] averaged over 1024~ms and for energies
between 50 and 300 keV.)
Consider a lensed burst from a source at redshift $z$,
with apparent {\em unlensed} brightness $b$.
What is the conditional probability that both images are
then brighter than the threshold $b_{\rm min}$?

The magnification $\mu_{-}$
of the fainter of the two images is $\mu_{-}=\left({r_{\rm E}/r_s}\right)-1$,
where $r_{\rm E}$ is the Einstein radius and $r_s$ the impact parameter
(see, e.g., \cite{GN94}).
Thus the fainter image will be brighter than $b_{\rm min}$ only if
$r_s/r_{\rm E} < \left(1+b_{\rm min}/b\right)^{-1}$,
so that the conditional probability we seek is 
$\left(1+b_{\rm min}/b\right)^{-2}$.
With $F_z(b)$ the fraction of bursts at redshift $z$ with apparent
unlensed brightness brighter than $b$,
the expected number $N$ of {\em observable} image pairs is now:

\begin{equation}
N = \epsilon N_{\rm tot} \int_0^\infty dC(z)\, \tau(z)
		\int_0^\infty dF_z(b)\, {1\over [1+ b_{\rm min}/b(z)]^2}  \; .
\end{equation}

Since the distribution of burst source redshifts, $C(z)$, is unknown,
we compute the expected number $N_{\langle z\rangle}$ 
of observable image pairs as a function of $\langle z\rangle$, 
the {\em effective} average redshift,
imagining that all sources are at this redshift:

\begin{equation}
N_{\langle z\rangle} = \epsilon N_{\rm tot} \tau(\langle z\rangle)
		\int_0^\infty dF(b)\, {1\over (1+ b_{\rm min}/b)^2}  \; .
\end{equation}
Equating the two previous equations serves as a definition for 
the effective redshift $\langle z\rangle$ of the burst distribution, 
and parametrizes our ignorance of the burst source distribution $C(z)$.
We take the unlensed brightness distribution, $F(b)$, to be the
observed BATSE brightness distribution,
since {\em de facto} no burst lensing
has been found. 

We define the integral in equation (5) to be $f(b_{\rm min})$,
the fraction of image pairs having both images
above the lightcurve--comparison brightness threshold, $b_{\rm min}$.
This fraction is the conditional probability
$\left(1+b_{\rm min}/b\right)^{-2}$
that both images are brighter than the threshold, 
weighted over the actual brightness distribution, $F(b)$, 
for the entire BATSE sample.
Figure~3 shows $f(b_{\rm min})$ as a function of $b_{\rm min}$
(in units of peak flux),
computed using the peak flux distribution $F(b)$ 
from the BATSE 4B catalog (\cite{M98}).

In estimating the effective feasibility of lightcurve comparison of
image pair candidates, previous investigators
have found that as many as 90\% of GRBs
are bright enough for such a task (\cite{NG94}).
This would imply that $b_{\rm min} = 0.27$ ph~cm$^{-2}$~s$^{-1}$
(\cite{M98}), and that $f(b_{\rm min}) = 0.57$.
In this work, we take a more conservative (and round) value of
$f(b_{\rm min}) = 0.5$.
This requires only that the brightest 80\%
of GRBs have lightcurves that can effectively be compared
for lens candidate identification,
a condition that is likely to be satisfied (see below).

\subsection{Results and upper limits}

Figure~4 shows the expected number, $N_{\langle z\rangle}$, 
of observable image pairs in the BATSE 4B catalog,
as a function of the effective average redshift, $\langle z\rangle$,
of GRB sources in the catalog.
We have used equation (5) and $\tau(z)$ from \S\ 2,
and have taken $F$=0.1 and $f(b_{\rm min})$=0.5.
We show our results for several different cosmologies.

Because $\tau(z)$ is a monotonically increasing function of $z$ 
(see Figs. 1 and 2), $N_{\langle z\rangle}$ is also a
monotonically increasing function of $\langle z\rangle$.
Hence, if there are no multiply imaged GRB sources
in the BATSE 4B catalog, there is
a corresponding upper limit to $\langle z\rangle$. 

To date, Marani \etal\ (1998) and Marani (1998)
have compared the lightcurves of the brightest 
75\% of the first 1235 BATSE bursts
and find no match among any of them.
Thus there is no evidence of gravitational lensing in these bursts.
The limits for $\langle z\rangle$ that we quote in this paper
consider the 1802 bursts in the BATSE 4B catalog, assuming that
80\% of them are bright enough for analysis and none of those
80\% are multiply imaged.
The heavy dashed horizontal lines in Figure~4 indicate the 68\% (lower
line) and 95\% (upper line) confidence limits to the number of lensing
events allowed in the BATSE 4B catalog.

The intersection points for these confidence limits are given in Table~1,
giving the 68\% and 95\% confidence-level upper limits on the
effective average redshift $\langle z \rangle$ for several
different cosmologies.
The strongest limits arise in cosmologies with a large
cosmological constant, where the lensing rate is quite high:
At the 95\% confidence level, we find that \hbox{$\langle z\rangle<2.2$}
if $\Lambda=0.7$ and $\Omega=0.3$.
The weakest limit arises in an Einstein--de Sitter universe,
where the lensing rate is smallest:
At the 95\% confidence level, we find that 
\hbox{$\langle z\rangle<5.3$}.

In Table~1 we also give the limits that result
from the current analysis of 1235 bursts (Marani \etal\ 1998; Marani 1998),
and those that would arise
if no lensing events are seen in a sample twice the current size, 
namely 3600 bursts.
Such a sample can be produced if BATSE
continues to operate in its current fashion for another five years.
Note that the resultant limits are much smaller
and less dependent on the cosmological parameters,
so that the continued increase in the number of GRBs observed by BATSE
will greatly constrain their redshift distribution (see \S\ 4).

Assuming no lensing in the current BATSE sample,
we find that the 68\% upper limit on $\langle z\rangle$,
as a function of the cosmological parameters,
is approximately given by:

\begin{equation}
\langle z\rangle < 1.5 + (\Omega - \Lambda) \pm 0.1  \; .
\end{equation}
This confirms the quasi--degeneracy in the lensing rate
(see, e.g., Kochanek 1993b) as a function of the cosmological
parameters, so that our upper limits are dependent
on the difference $\Omega -\Lambda$. Studies of 
high--redshift supernovae (see, e.g., Perlmutter et al. 1997, 1998;
Garnavich \etal\ 1998) also constrain $\Omega-\Lambda$: 
This is not surprising, since
their results are also based upon the redshift--distance relation.
Riess \etal\ (1998) find a best--fit value for
$\Omega-\Lambda$ of --0.5, with a systematic error of
approximately $\pm 0.3$. Systematic errors arising from
reddening corrections could in principle allow the supernova
data to be consistent with \hbox{$\Omega-\Lambda$} as large
as +0.4, but preliminary indications are that these
corrections are small (A. Riess, private communication).
Hence the 68\% upper limit on $\langle z\rangle$ is less than 1.9,
and is probably close to 1.0.
Current star-formation models, for which  $\langle z\rangle\sim 2.2$,
are marginally inconsistent (1 $\sigma$) with our limits,
regardless of the cosmology.
If $\Omega-\Lambda < -0.4$, these models are inconsistent 
with our limits at the 2~$\sigma$ level.

These upper limits to $\langle z\rangle$, combined with the
median fluence of $10^{-5}$erg cm$^{-2}$ of the BATSE
bursts (see Meegan et al.\ 1998) imply that, at the 68\%
confidence level, the upper limit to the median rest-frame 
energy release of the BATSE gamma-ray bursts is between
$1.8\times 10^{52}h^{-2}$ ergs (for \hbox{$\Omega=0.3$} and
\hbox{$\Lambda=0.7$}) and $2.9\times 10^{52}h^{-2}$ ergs 
(for \hbox{$\Omega=1$} and \hbox{$\Lambda=0$}) for the
energy observed in the BATSE window, where
the energy is assumed to be emitted into $4\pi$ steradians.

\section{DISCUSSION AND SUMMARY}

We have exhibited a new method for calculating gravitational lensing
rates, given a cosmology and a distribution and evolution of the lenses.
This method, which is described more fully in Holz \& Wald
(1998), is free of the angular diameter distance ambiguity that is
a feature of standard analytical methods. For the cosmologies and
redshift ranges we have examined, we find that the lensing rate is
approximately equal to the filled-beam lensing rate for $z<3$, but
in excess of the filled-beam rate for $z>3$. We find that the lack
of detected lensing in the current BATSE catalog places an upper
limit to the median redshift of bursts that, at the 68\% confidence
level, is comparable to or less than the median redshift in
star-formation burst models.

We now discuss these results and place them in context. In
\S~4.1 we discuss the simplifying assumptions that have been  made in our
calculation. We find that these assumptions have, if anything, caused us
to {\it underestimate} the lensing rate. In \S~4.2 we examine the
uncertainties in the inputs used in our calculations, as well as the
future prospects for reducing these uncertainties. In \S~4.3 we discuss
the implications of the current lack of lensing for cosmological models
of gamma-ray bursts, when combined with existing lower bounds to the
redshift of bursts in these models. Finally, in \S~4.4 we provide a
future outlook for the rapidly emerging importance of constraints
inferred from detection or nondetection of lensing.

\subsection{Effects of Simplifying Assumptions}

We have assumed that the angular distribution of the sources of gamma-ray
bursts is uncorrelated with the angular distribution of the lenses. This
is well justified, because for sources at low redshift the lenses that
contribute most to the lensing rate are at about half the redshift to
the source, and for sources at high redshift the dominant lenses are at a
redshift of approximately unity (see \S~2; see also, e.g., Turner et al.\
1984; \cite{M92}). 
Hence, in a cosmological model the sources of gamma-ray bursts
are separated from lenses by hundreds of megaparsecs to gigaparsecs, and
therefore it is extremely unlikely that the sources and lenses are
angularly correlated. We have also assumed that the phase of the orbit of
BATSE is uncorrelated between the separate images of the burst, and hence
that the joint probability of two images being observable is just the
square of the BATSE efficiency, $0.48^2 \approx 0.23$. This assumption is
reasonable if the time delay between images is significantly larger than
the BATSE orbit time of $\sim 5000$ seconds. The time delay for a mass
$M$ is of order $10^{-5}(M/M_\odot)$ seconds (see, e.g., Blandford \&
Narayan 1992), and hence the assumption of uncorrelated phases is good
for masses $M\gta 10^9\,M_\odot$. This includes almost all the effective 
lensing mass of galaxies, and thus this assumption is also unlikely to
affect the calculated lensing rates significantly.

We may have underestimated the lensing rate by assuming that only
galaxies---which we have modeled as singular
isothermal spheres---contribute to the lensing rate.
Nemiroff et al.\ (1993) have noted that point
masses, for example supermassive black holes, could also in principle 
contribute. To produce gamma-ray burst lensing of the type that
we consider here, in which the lensing event appears as two
separate bursts, the mass of such a lensing black hole must be
$M\gta 10^8\,M_\odot$, because smaller masses would produce
time delays less than $\sim$1000 seconds, so that overwriting
or readout time would tend to prevent BATSE from detecting two
separate bursts (the effect of lensing by lower-mass black 
holes may, however, be detectable using techniques such as
autocorrelation analysis; see Nemiroff et al.\ 1994, 1998).
If, however, the total lensing rate were dominated by
very massive black holes, $M\gta 10^{10}\,M_\odot$, then these
objects would also produce detectable image separation of
lensed quasars or galaxies 
(the angular separation is of order $3(M/M_\odot)^{1/2} (D/1
{\rm Gpc})^{-1/2}\,\mu$arcsec for an Einstein ring, which is 0.3" for
$M=10^{10}\,M_\odot$ and $D$=1 Gpc). 
Hence, the amount by which we have underestimated the lensing
rate depends on the number of point masses in the relatively narrow range
$10^8$--$10^{10}\,M_\odot$, which is the only mass range that
could avoid detection in quasar
lensing surveys and yet produce separately detected bursts.

We may also have underestimated the lensing rate by 
only keeping track of the total number of images, not whether there are,
e.g., two or four images in a particular event. As pointed out by
Grossman \& Nowak (1994), this tends to underestimate the {\it
detectable} rate of lensing because when there are several images it is
more probable that at least two of them are observed by BATSE: assuming a
48\% efficiency for any particular image, the probability of observing at
least one pair rises from 23\% when there are two images to 66\% when
there are four images. Grossman \& Nowak (1994) estimate that, for galaxy
ellipticities typical of galaxy surveys, the maximum possible
enhancement of the lensing detection rate is $\sim$30\%, but that the
overall enhancement is likely to be smaller.

A third effect that may enhance the rate of lensing is magnification
bias, which is an effect first discussed extensively 
by Fukugita \& Turner (1991) in the context of quasar lensing.
Magnification bias occurs because sources that would have been
undetectable are made visible by lensing, and hence a 
magnitude-limited sample contains an enhanced incidence of
lensing. This can be especially important if the number of
sources at a given flux rises steeply at the faint end, as
it does in quasars. To detect multiple images and thus confirm
strong lensing, it is necessary to detect the fainter image
as well as the stronger image. For persistent sources such as
quasars, this can be done by following up broad, shallow surveys
by deep pointings, so that essentially all of the secondary
images are detectable (e.g., as in the Hubble Snapshot Survey
[Bahcall et al. 1992; Maoz \etal\ 1992, 1993a,b]). 
For transient sources such as gamma-ray
bursts, deep follow-ups are not always possible, and hence for
the secondary image to be detectable it must have a brightness
in excess of the threshold of the original survey (i.e., in the
case of gamma-ray bursts, the BATSE sample). This effect, plus
the flatness of the faint end of the gamma-ray
burst log N -- log P curve, suggests that the magnification bias for
gamma-ray bursts is probably less than the magnification bias for quasars
(see also \cite{GN94} for a discussion of this point). Nonetheless,
magnification bias for gamma-ray bursts could in principle
increase significantly the expected lensing rate, and hence decrease 
the upper limits on $\langle z\rangle$, compared to the
conservative estimates here.

We may underestimate the lensing rate because we have
neglected the clustering of galaxies, the last of our assumptions. 
The enhanced gravitational potential in clusters tends to increase 
the convergence of
null geodesics and hence increase the cross section for strong lensing
(see also Holz \& Wald 1998). Numerical estimates suggest that the
lensing rate for galaxies in a cluster will be increased by
$\sim$10--20\% by this effect, but because only $\sim$10\% of galaxies
are in clusters this effect is unlikely to increase the overall lensing
rate significantly.

Finally, however, we may have overestimated the rate of
lensing by ignoring evolution in the properties of the lensing galaxies.
If the comoving density in lensing galaxies was less in the 
past than it is now, or if these galaxies were less massive than they
are today, high-redshift lenses would
contribute less to the lensing rate than we have assumed. Note,
however, that the peak in the lens redshift distribution is at about
unity (Fig.~2), when galaxies had properties very similar to their current
properties. Note also that observations of lensing of quasars (see
below), which have a typical redshift similar to the
proposed typical redshift $z\sim 2$ of GRBs, suggest that if
there are evolutionary effects then these effects have not had a
dominant effect on the lensing rate.

To summarize, the net effect of our simplifying assumptions is likely to
be that we underestimated the lensing rate by $\lta$10\%, and hence our
results our conservative in that they give a slightly high upper bound to
$\langle z\rangle$.

\subsection{Uncertainties in Input Parameters}

In addition to the simplifying assumptions discussed above, our
calculations are clearly dependent on the input values we have
assumed. One input is the dimensionless parameter $F$, which is
proportional to the product of the number density of galaxies and the
fourth power of their average velocity dispersion. The lensing rate
scales linearly with $F$.  We have used $F$=0.1, which is consistent with
the recent estimates from the Century Survey (Geller et al.\ 1997). Other
estimates range from $F$=0.05 to $F$=0.15, and
if these other estimates are used then the range of possible
maximum redshifts is increased for a given cosmology. Fundamental
cosmological parameters, another input to our calculations, are also
uncertain. A standard Einstein-De Sitter universe ($\Omega$=1,
$\Lambda$=0) gives the lowest lensing rate, and a $\Lambda$-dominated
universe gives the highest.

Current observation of quasar lensing puts constraints on the allowed
combinations of $F$ and cosmology. For example, five of the 351
quasars in the HST Snapshot Survey (Maoz et al.\ 1993b) were lensed.
A comparison of this rate with the rate expected for models is
complicated by effects such as magnification bias (see above).
However, the high average redshift of the quasars ($z\sim 2$) means
that many of the issues affecting lensing of gamma-ray bursts,
such as evolution of clustering, also affect the lensing of quasars.
Hence, quasar lensing statistics have bearing on estimated GRB
lensing rates, and will become even more relevant as more extensive
surveys are done. 

In addition, all of these uncertainties will be diminished greatly by the
data that emerge from the plethora of satellites and surveys planned for
the next few years. The number density and velocity distribution of
galaxies (and hence $F$) will be determined with unprecedented accuracy
by surveys such as 2dF (Colless 1998) 
and the Sloan Digital Sky Survey (Margon 1998), 
the latter of which is expected to see first light in 1998. 
Data from these surveys will also provide
valuable information about  $\Omega$, $\Lambda$, clustering, and the
lensing rate of quasars. Complementary information about cosmological 
parameters will be extracted
from high-redshift supernova surveys (Perlmutter \etal\ 1998) and from data 
gathered by the many upcoming microwave background
experiments, such as MAP (Wang, Spergel, \& Strauss 1998), 
TOPHAT (Martin et al.\ 1996), 
and PLANCK (see, e.g., Bond, Efstathiou, \& Tegmark 1997 
for a discussion of the constraints).
The evolution of galaxy clustering is already being inferred from
cluster surveys, and AXAF observations of X-ray
emission from clusters of galaxies is expected to greatly enhance
this understanding.
Hence, one side
effect of the coming data-rich era in cosmology is that calculations
of lensing rates will be much less uncertain, and
therefore upper limits to the redshift of a population such as gamma-ray
bursts will be far more secure.

\subsection{Current Implications for Cosmological GRB Models}

The redshift limits presented in this paper
have important consequences for cosmological models of GRBs.
Lower limits on the redshift of bright GRBs are becoming stronger as
the number of detected bursts rises, both because of the no-host problem
(Schaefer et al.\ 1997) and because GRBs do not show any
evidence of large-scale clustering (Lamb \& Quashnock 1993; Quashnock 1996). 
These limits suggest that if the sources of GRBs are in or near galaxies, 
the median redshift of the bursts detected with BATSE 
may be significantly greater than unity. 
However, the upper limits to the redshift  of GRBs that
follow from the lack of lensing are beginning to make such  high-redshift
populations less appealing. For example, the lack of lensing in the
current BATSE catalog strongly rules out the \hbox{$z\sim 10$--200}
models that were heretofore consistent with the properties of the
BATSE bursts (Rutledge et al.\ 1995). Also, models in which the burst rate
is proportional to the star formation rate (Totani 1997;
Wijers et al.\ 1998), and for which $\langle z\rangle\sim 2.2$,
are beginning to be constrained strongly 
by the lack of lensing: Our 68\% upper limit on $\langle z\rangle$
(eq.~[6]) is less than 1.9, and probably close to 1.0.
If the redshift upper limits are reduced further, 
it may be necessary to postulate either a very short interval in
which GRBs were produced, with an intrinsic brightness distribution that
matches the observed log N -- log S distribution, or a population that is
unassociated with any known population of objects, so that the lower
limits to redshift do not apply.

\subsection{Future Outlook and Summary}

In the next few years gravitational lensing is likely to play a role  of
rapidly increasing importance in constraining cosmological models of
gamma-ray bursts. As discussed in \S~4.2, the uncertainties in the
calculation of the expected lensing rate will be diminished greatly, and
hence the limits will be robust. In addition, as BATSE continues to
detect bursts the statistics will steadily improve, and in the
current high-redshift models lensing is to be expected
in the next few years. In five years, the
sample will be roughly double the $\sim$1800 bursts considered here, and
therefore if lensing continues to be absent, upper limits on the redshift will 
become extremely restrictive. At the same time, the improved statistics
will continue to increase the lower limits on redshift, if there is no
apparent large-scale clustering (Lamb \& Quashnock 1993; Quashnock 1996).
Moreover, the accurate positional estimates of BeppoSAX,  HETE II, and
the Fourth Interplanetary Network (IPN) are likely to provide $\sim$100 error
boxes of area a few square arcminutes or less in the next five years.

If instead lensing {\it is} detected, there will be
many important  consequences. For one, the cosmological origin of a
particular GRB will be established from gamma-ray data for the first time
(as opposed to from the afterglow, as may be the case for GRB~970508
[\cite{M97}]). The approximate mass of the lens, and hence its 
luminosity, can be estimated from the time delay between images. The
sharp peak in the probable redshift of lenses (see \S~2) would then give
a reasonably accurate estimate of the flux of the lens. If the GRB can be
reasonably well-localized (e.g., with the IPN),
this will allow a search for the lens. 
If the lens is found, the GRB source must
essentially be angularly coincident with it, because the deflection angle
is only about 1", and thus counterpart searches will be facilitated
greatly. Knowledge of the redshift of the lens, combined with $H_0$ and an
independent estimate of the mass of the lens (e.g., from its luminosity
or velocity dispersion) will allow use of the time delay to estimate the
redshift of the source. 
The quantifiable diversity of gamma-ray burst light curves 
(\cite{MN98}; \cite{Mar98}) means that false positives are unlikely, 
and hence even a single lensing event will generate a wealth of data.

In conclusion, the increased estimates of the redshift of gamma-ray
bursts that were forced by the no-host problem, in addition to the
greatly improved statistics of bursts and
the increase in the estimate of the BATSE burst detection
efficiency from 34\% to 48\%,
mean that gravitational lensing
limits are now playing a prominent role in constraining gamma-ray burst
models. The role of lensing will become dramatically more important and
robust in the next five years.

\acknowledgements

We thank Don Lamb for a careful reading of the manuscript
and for discussions of the importance of multiple
encounters.  We also acknowledge valuable
discussions with Bob Wald and Asantha Cooray.
We thank the referee, Bob Nemiroff, for a number of helpful comments.
This work was supported in part by NSF grant PHY~95-14726 (DEH),
by NASA grant NAG~5-2868 (MCM), 
and by NASA grant NAG~5-4406 and NSF grant DMS~97-09696 (JMQ).

\newpage
\begin{figure}
\figurenum{1}
\plotone{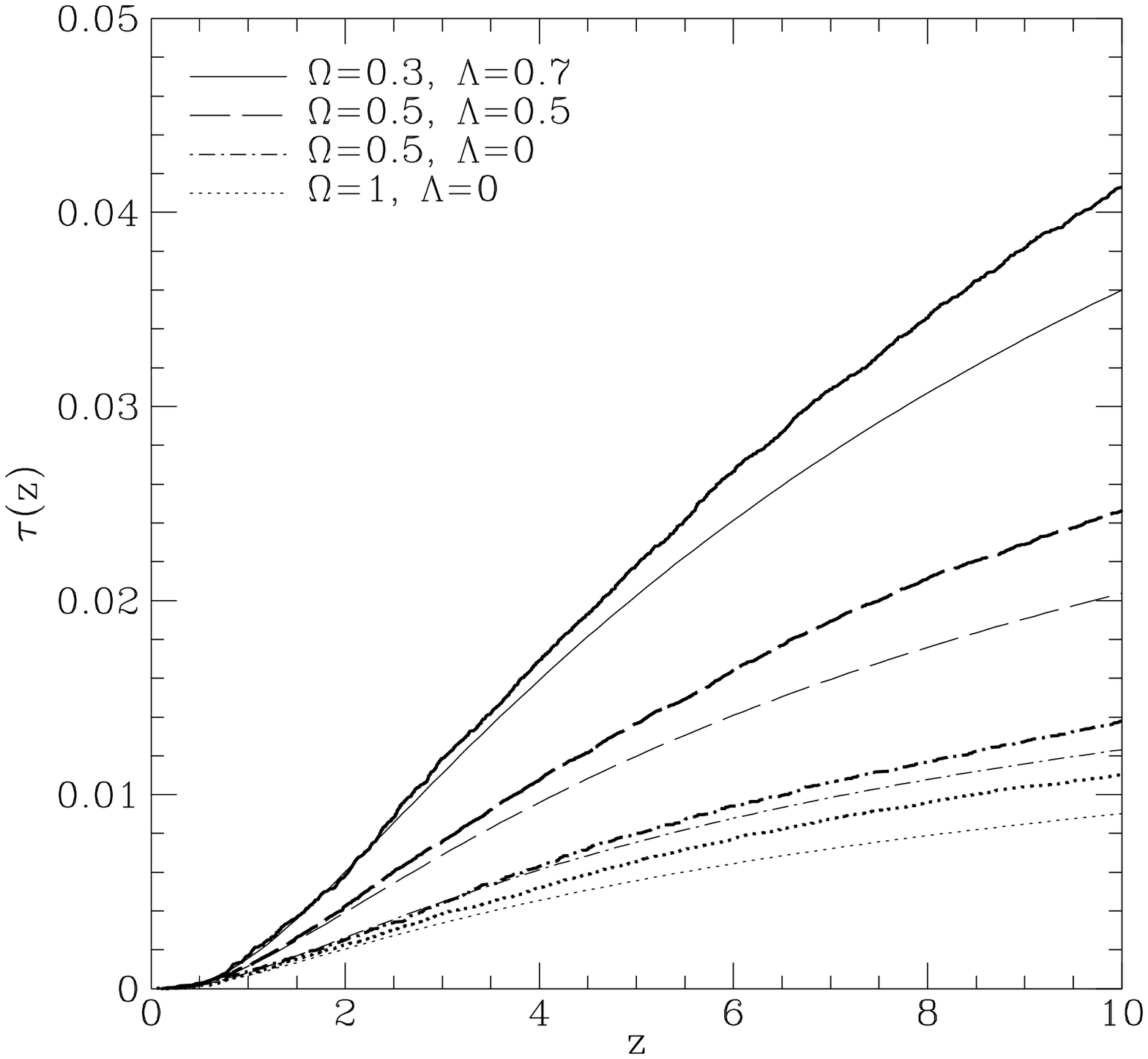}
\caption{Lensing probability, $\tau(z)$, 
as a function of redshift, $z$,
for a range of different cosmologies.
The lenses are singular isothermal spheres, and
the dimensionless parameter $F=0.1$ (see text). 
The values of $\Omega$ and $\Lambda$ for each curve
are indicated by the legend. The boldface curves are the numerical
lensing rates calculated using the procedure described in \S~2.
The light curves are the analytic rates calculated using the
filled-beam approximation, which yields curves much closer to the numerical
rates than the empty-beam (\cite{DR}) 
or Ehlers-Schneider (\cite{ES})
prescriptions. \label{fig1}}
\end{figure}

\newpage
\begin{figure}
\figurenum{2}
\plotone{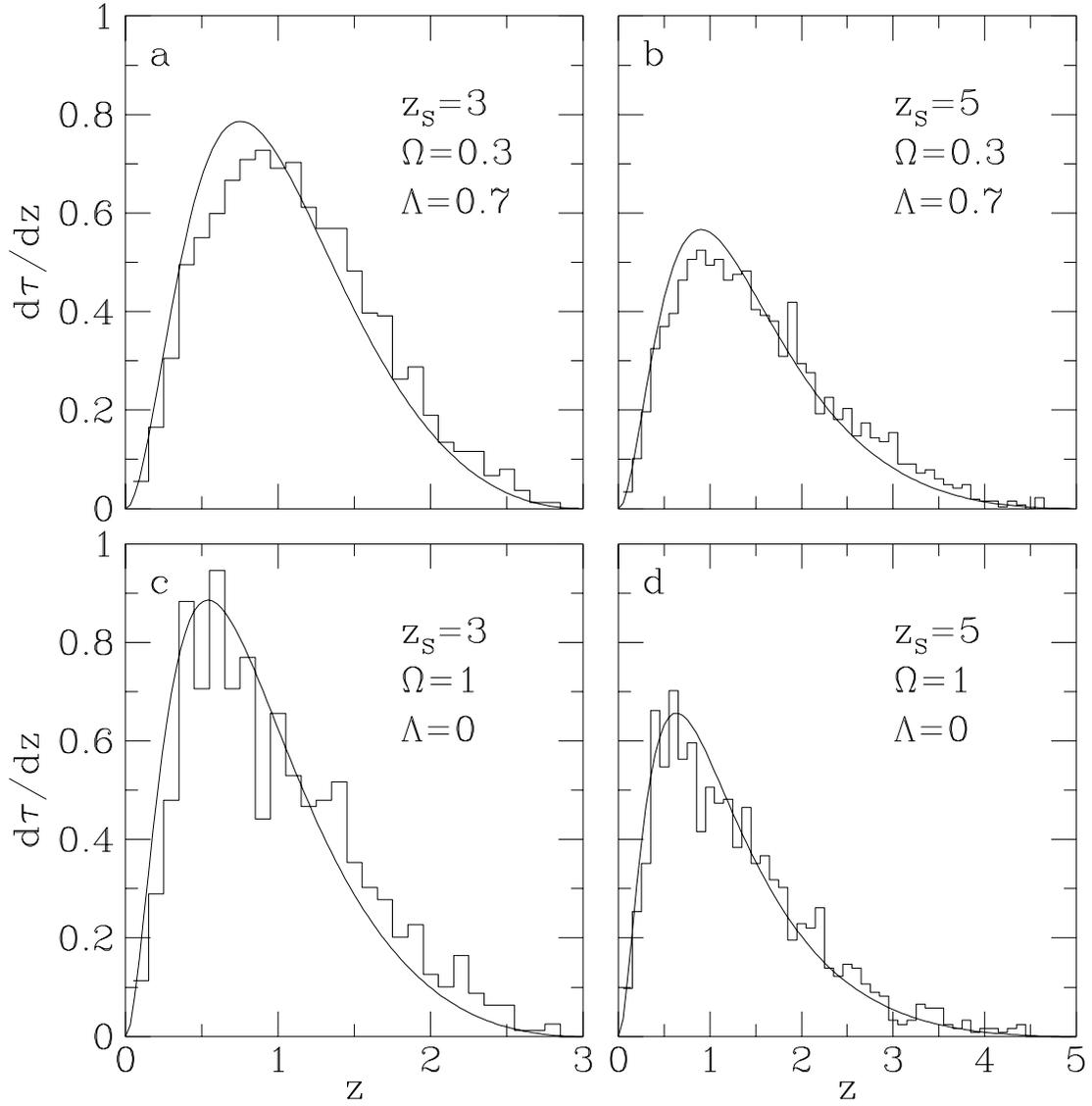}
\caption{Differential redshift distribution of lenses, 
$d\tau/dz$, for $\Omega$=0.3 and $\Lambda$=0.7 and a source redshift of
$z_S$=3 ({\em panel} a) or $z_S$=5 ({\em panel b}), 
and for $\Omega$=1 and $\Lambda$=0 with $z_S$=3 ({\em panel} c)
or $z_S$=5 ({\em panel} d). 
The histograms give the numerical distribution, 
and the solid curves give the analytic distribution
in the filled-beam approximation (Turner \etal\ 1984).
The lenses are singular isothermal spheres. \label{fig2}}
\end{figure}

\newpage
\begin{figure}
\figurenum{3}
\plotone{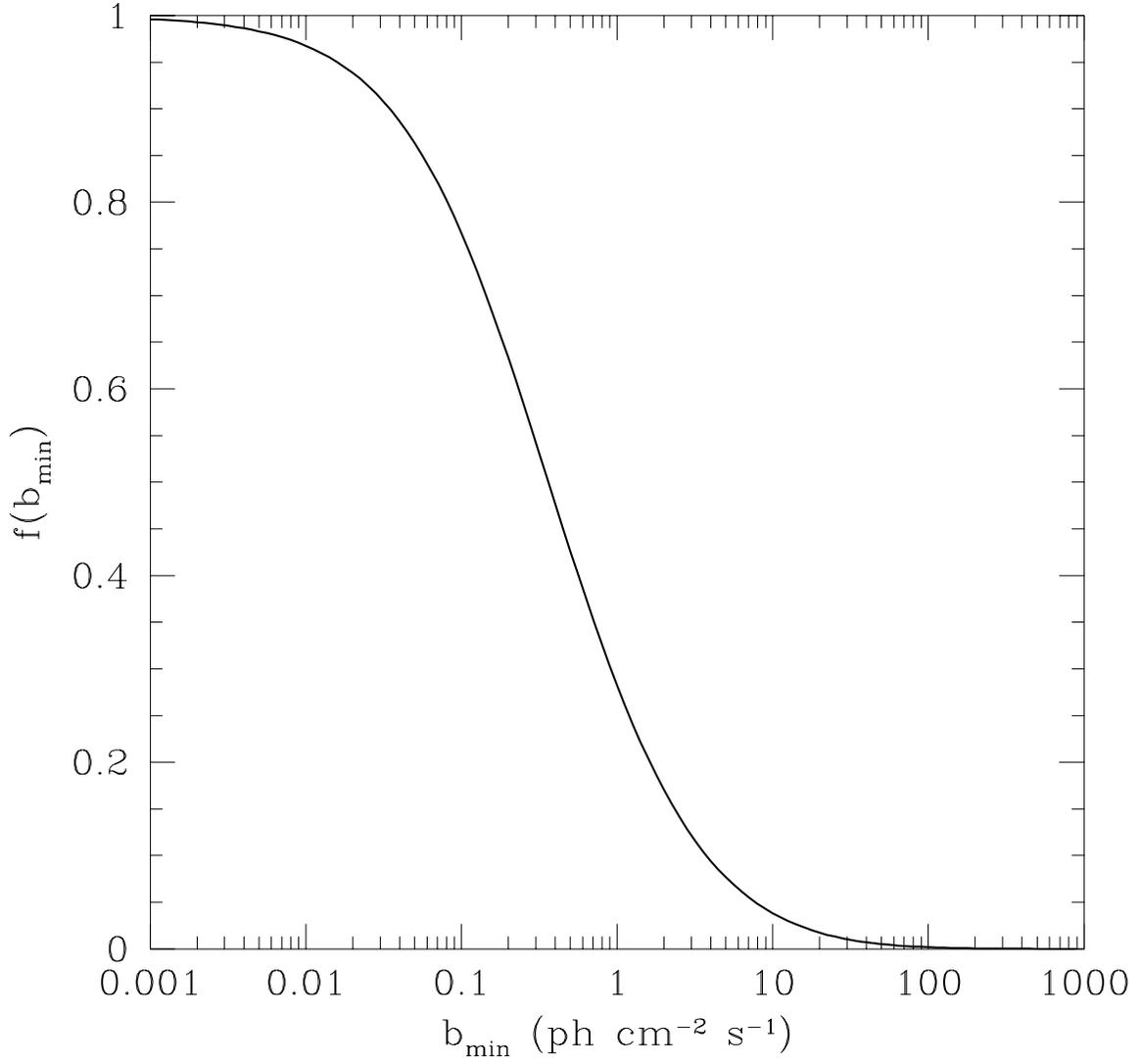}
\caption{The fraction of image pairs, $f(b_{\rm min})$,
having both images above the lightcurve--comparison brightness threshold, 
$b_{\rm min}$ (in units of peak flux averaged over 1024~ms). \label{fig3}}
\end{figure}

\newpage
\begin{figure}
\figurenum{4}
\plotone{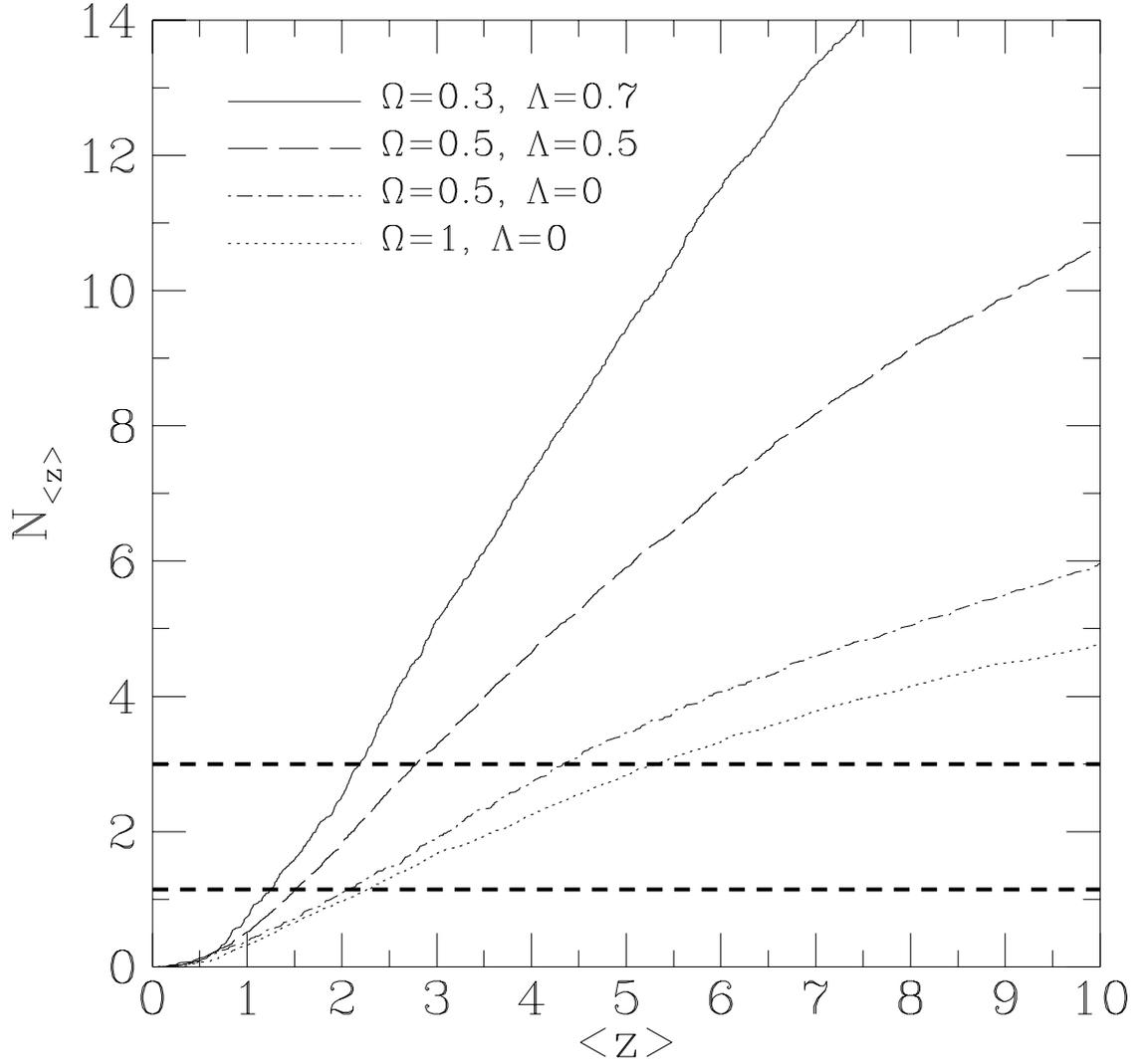}
\caption{Expected number, $N_{\langle z\rangle}$, 
of observable image pairs in the BATSE 4B catalog, 
as a function of $\langle z\rangle$, 
the {\em effective} average redshift of GRB sources in the catalog,
assuming that $F$=0.1 and $f(b_{\rm min})$=0.5.
The values of $\Omega$ and $\Lambda$ for each cosmology
are indicated by the legend.
The heavy dashed horizontal lines indicate
the lensing rates that are discrepant with the
lack of detected lensing at the 68\% (lower line) and
95\% (upper line) level. \label{fig4}}
\end{figure}

\newpage

\begin{deluxetable}{ccccc}
\tablewidth{500pt}

\tablecaption{
 \label{table:zlimit}
 Limits on the average redshift $\langle z\rangle$ for different cosmologies}

\tablehead{Confidence Level&$\Omega$=0.3, $\Lambda$=0.7&
$\Omega$=0.5, $\Lambda$=0.5&$\Omega$=0.5, $\Lambda$=0.0&
$\Omega$=1, $\Lambda$=0}

\startdata

Current analysis (1235 bursts)\\
\\
95\%&2.7&3.8&6.6&8.6\\
68\%&1.5&1.9&2.7&3.0\\
\\
\\
Current data (1802 bursts)\\
\\
95\%&2.2&2.8&4.3&5.3\\
68\%&1.2&1.5&2.1&2.3\\
\\
\\
3600 bursts with no lensing\\
\\
95\%&1.4&1.8&2.6&2.8\\
68\%&0.9&1.1&1.3&1.4\\

\enddata

\end{deluxetable}

\end{document}